\documentclass[useAMS,usenatbib]{mn2e}
\usepackage{graphicx}

\def\V{{\mathcal V}}
\def\Sr{S_r}
\def\Rsol{R$_\odot$}
\def\Msol{M$_\odot$}

\title[Exoplanet transit detection with red noise]{The effect of red noise on  planetary transit detection}
\author[Pont, Zucker and Queloz]{Fr\'ed\'eric Pont$^{1}$\thanks{E-mail:
frederic.pont@obs.unige.ch}, Shay Zucker$^{2,3}$ and Didier Queloz$^{1}$\\
$^{1}$Geneva University Observatory, 1290 Sauverny, Switzerland\\
$^{2}$Faculty of Physics, Weizmann Institute of Science, P.O. Box 26,
Rehovot 76100, Israel\\
$^{3}$Present address: Department of Geophysics and Planetary Sciences, Beverly and Raymond
Sackler Faculty of Exact Sciences, Tel Aviv University, Tel Aviv 69978, Israel}

\begin{document}

\bibliographystyle{mn2e}

\date{Accepted  . Received  }

\pagerange{\pageref{firstpage}--\pageref{lastpage}} \pubyear{2006}

\maketitle

\label{firstpage}

\begin{abstract}

Since the discovery of short-period exoplanets a decade ago,
photometric surveys have been recognized as a feasible method to
detect transiting hot Jupiters. Many transit surveys are
now under way, with instruments ranging from $10$-cm cameras to the
Hubble Space Telescope. However, the results of these surveys have
been much below the expected capacity, estimated in the
dozens of detections per year.

One of the reasons is the presence of systematics (``red noise'') in
photometric time series. In general, yield
predictions assume uncorrelated noise (``white noise''). In this paper, we
show that the effect of red noise on the detection threshold and the
expected yields cannot be neglected in typical ground-based surveys.
We develop a simple method to determine the effect of
red noise on photometric planetary transit detections. This method can
be applied to determine detection thresholds for transit
surveys. We show that the detection threshold in the
presence of systematics can be much higher than with the assumption of
white noise, and obeys a different dependence on magnitude,
orbital period and the parameters of the survey. Our method can also
be used to estimate the significance level of a planetary transit
candidate (to select promising candidates for spectroscopic
follow-up).

We apply our method to the OGLE planetary transit search, and show
that it provides a reliable description of the actual detection
threshold with real correlated noise.  We point out in what way the
presence of red noise could be at least partly responsible for the
dearth of transiting planet detections from existing surveys, and
examine some possible adaptations in survey planning and strategy.
Finally, we estimate the photometric stability necessary to the
detection of transiting ``hot Neptunes''.
 

\end{abstract}


\section{Introduction}

Many photometric surveys for transiting exoplanets are now under way, with a wide variety of instrumentation and observation strategies -- from monitoring large areas of the sky with small telescopes to deeper surveys on star clusters or the Galactic disc with $1$--$4$m telescopes\footnote{see e.g. \citet{alo04,bak04,bra06,bru03,hid05,hoo05,mal03,rau04, uda02a} for a description of some planet transit surveys; \citet{cha06} for a recent review of the results.}. Altogether, thousands of telescope nights have been invested in these surveys, monitoring hundreds of thousands of target stars in the Solar neighbourhood and in the Galactic disc.  However, even after years of operation, the results of these surveys failed to meet the expectations, with only a slow trickle of detections instead of the expected bounty.

Part of the mismatch between expectations and actual performance can
be attributed to the fact that these surveys were often assuming that 
the first known transiting extrasolar planet and for
long the only one, HD209458$b$ \citep{cha00}, was typical of hot
Jupiters. Its radius of $\sim1.3$~R$_J$ creates a $2\%$ transit signal
on a solar-type star. Subsequently discovered transiting gas giants
showed the large radius of HD209458 to be an exception rather than the
rule \citep{alo04,pon04}. The mode of hot Jupiter radii is probably
near $1.1$~R$_J$ or lower \citep{gau05}, producing correspondingly shallower eclipses.

Still, even assuming smaller gas giants, transit surveys keep
predicting many more detections than they actually yield.
Understanding this mismatch is essential both for interpreting the
results in terms of planetary statistical properties, and in order to
improve the planning and strategy of the surveys. Indeed, if the
statistical properties of hot Jupiters were not known from radial
velocity surveys, it is very likely that drastically different
conclusions would have been drawn from the results of the transit
surveys. Upper limits would have been put to the abundance of hot
Jupiters one or two orders of magnitude below those derived from
Doppler surveys.

The ingredients in the prediction simulations seem solid: the
abundance of hot Jupiters is relatively well known from radial
velocity surveys, the statistics of targets are obtained from well-tested
models of Galactic stellar populations, and the rest comes from simple
orbital mechanics. The setting of the transit detection threshold is
often considered a minor component in the simulations. The threshold
is generally modelled as a minimum signal-to-noise ratio (``S/N'') of
the transit detections assuming uncorrelated noise in the photometric
data.

In this paper we show that, contrary to these assumptions, the correlation of photometric data at the millimagnitude level cannot be neglected when determining the detectability of planetary transits, and that taking the correlation into account can strongly affect the detection threshold, and consequently the estimates of the potential of photometric surveys in terms of planet detetion.

We develop a simple method to assess the significance of detected
transit candidates in the presence of such ``red noise''. We propose a
method that is robust and gives realistic results, while at the same
time remains simple to use and to apply to any ground-based transit
survey.

Ongoing transit surveys have shown that detecting transiting signals
was not the end of the story. For transit depths in the range of a few
percents, by far the largest number of detections are due to eclipsing
binaries, either small transiting M dwarfs \citep[e.g. ][]{pon05} or
eclipsing binaries diluted by the light of an unresolved companion
\citep[e.g. ][]{man05}. The identification of true transiting planets
among all identified transit candidates requires a considerable
investment in spectroscopic follow-up observations. For transit depths
below $\sim 2\%$, the odds become more favourable to transiting
planets \citep{bro03}, but since such transit depths are near the
detection threshold of most ground-based surveys, false positive
detections start being a source of contamination. False positives
require even more follow-up observations than eclipsing
binaries. Therefore the assignment of reliable significance levels to
transit detections near the detection threshold, is another motivation
for a robust method to assess the significance of transit candidates
in the presence of systematics or ``red noise''.

A third motivation is to derive realistic estimates for the uncertainty on planetary parameters derived from transit lightcurves, taking into account the effect of systematics in the photometry. 

In Section~\ref{Sec2} we present our method to compute the
significance level of a transit detection. In Section 3 we examine
the implications for transit surveys. Section 4 summarizes the results
of this paper and points out some interesting consequences.

\section{Modified detection statistic for red noise}

\label{Sec2}
\subsection{White-noise statistics}

\begin{figure}
\resizebox{8cm}{!}{\includegraphics{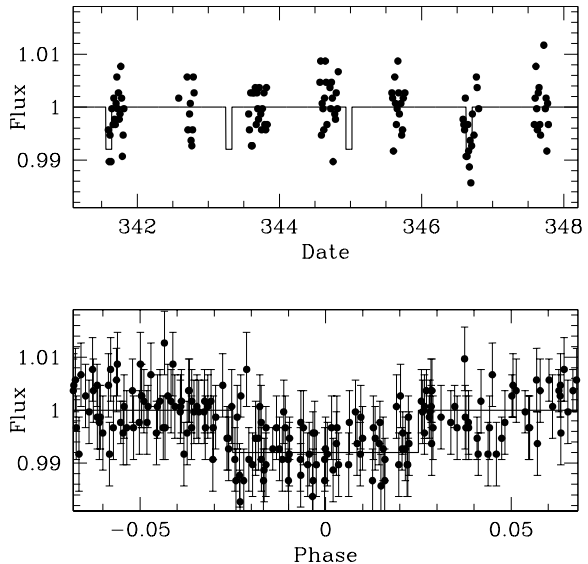}}
\caption{Lightcurve for a planetary transit candidate, with the step 
function used in the detection procedure
\citep[OGLE-TR-132,][]{uda03}. Top: short excerpt of the
lightcurve. Bottom: phase-folded lightcurve around the detected transit.}
\label{square}
\end{figure}

The signal produced in a stellar lightcurve by a planetary transit can be approximated by a strictly periodic step function (Fig.~\ref{square}), with a depth related to the radius ratio of the two bodies, and a duration related to the orbital elements and the primary radius. Transit detection algorithms usually work by fitting a step function to the phase-folded signal, or by detecting a step-like decrease and increase in the flux.

Let us assume that a possible transit signal has been detected in a
lightcurve consisting of $N$ flux measurements $f_i$, with
uncertainties $\sigma_i$. The flux is normalised so that the mean flux
outside the transit signal is $1$, $<f_i^{out}>=1$ .

For simplicity we first assume that the measurement uncertainties are equal for all data points, $\sigma_i \equiv \sigma_0$ (Section~\ref{app} discusses the generalisation to unequal uncertainties). Let $d$ be the best-fitting transit depth, and $n$ the number of data points in the transit. $d$ will then be the difference between the mean of the data points in the transit and the flux level outside the transit:
\[
d=  1 - <f^{in}> = 1 - \frac{\sum f_i^{in}}{n} \ .
\]

The uncertainty on $d$ is the error on the mean of $f_i^{in}$ (since $n<<N$ for planetary transits, we neglect the error on  $<\! f_i^{out}\! >$). Using the expression for the uncertainty on the mean under the assumption of uncorrelated noise gives:
\begin{equation}
\sigma_d = \sigma(<f_i^{in}>)= \sigma_0 / \sqrt{n} \ .
\label{sigd}
\end{equation}
The uncertainty on $d$ decreases with the square root of the number of points in the transit.

Intuitively, we can interpret $d$  as the ``signal'' in the transit detection procedure, and $\sigma_d$ as the noise. A natural statistic for the significance of a transit detection is thus the S/N ratio of $d$, which we write $S_d$:
\begin{equation}
S_d \equiv d/\sigma_d = \frac{d}{\sigma_0} n^{1/2} \ .
\label{white_snr}
\end{equation}
This expression is the one most commonly used to evaluate the significance of transit detections as well as to predict detection thresholds from the characteristics of the observations. Statisticians sometimes refer to $S_d$ as the 'Wald test' statistic.

\begin{figure}
\resizebox{8cm}{!}{\includegraphics{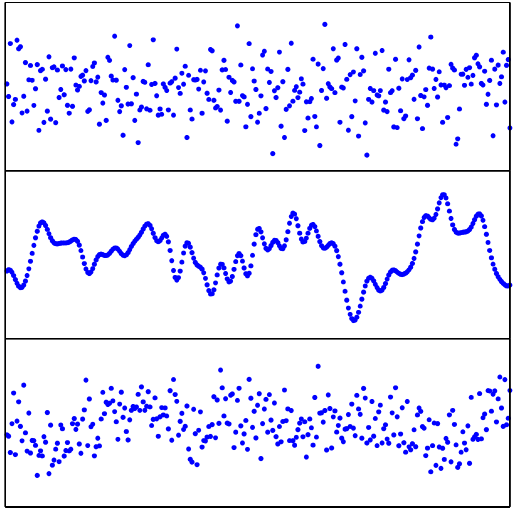}}
\caption{Lightcurves with white noise only (top), red noise only (middle) and white and red noise (bottom). Typical light curves from high-precision rapid time-series photometry for bright targets in transit surveys resemble portions of the bottom curve.}
\label{wrnoise}
\end{figure}

\subsection{Correlated residuals and coloured noise}

\begin{figure}
\resizebox{8cm}{!}{\includegraphics{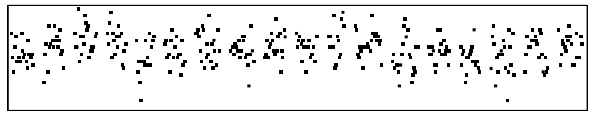}}
\caption{Example of a real lightcurve from a planetary transit survey (the OGLE survey). The intervals between different nights were compressed for the display. The vertical range of the plot is $0.03$~mag. Systematic trends are visible on several timescales to the level of a few millimagnitudes. }
\label{reallc}
\end{figure}

The derivation of Eq.~\ref{white_snr} requires a crucial assumption:
the photometric errors were assumed to be uncorrelated, i.e. purely
white noise was assumed. In fact, the errors on ground-based
millimagnitude photometry in rapid time series are correlated. Trends,
or ``systematics'' are present, related to changing airmass,
atmospheric conditions, telescope tracking, flatfield errors (usually
a combination of several of these factors). These effects introduce
some covariance between the lightcurve data points, with timescales
similar to the duration of planetary transits. Typical transits for
close-in planets last $2$--$3$ hours, which is also the timescale of
airmass, seeing and tracking changes in ground-based photometric
observations. In the parlance of signal analysis, the noise on
photometric observations is quite {\em red}, with a low-frequency
component.  Figure~\ref{wrnoise} illustrates the difference between
white noise and noise with a red component. Figure~\ref{reallc} gives
a real-life example of a lightcurve from a planetary transit
search. Red noise is apparent at the millimagnitude level in this time
series.

\subsection{Transit depth uncertainty with covariance}

The equivalent of Eq.~\ref{sigd} in the presence of correlated noise is:
\begin{equation}
\sigma_d^2 = \frac{1}{n^2}   \sum_{i,j}  C_{ij}= \frac{\sigma_0^2}{n} + \frac{1}{n^2} \sum_{i\neq j} C_{ij} \ ,
\label{sigcov}
\end{equation}
where $C_{ij}$ are the covariance coefficients 
between the $i^{th}$ and $j^{th}$ measurements, where the $i$ and $j$ indexes cover the measurements taken during the transit. The diagonal of the matrix $C$
contains the individual errors $\sigma^2_i$ (random uncertainty on the
$i^{th}$ measurement), and in Eq.~\ref{sigcov} they are all assumed equal to $\sigma_0$.

The first part of this expression is Eq.~\ref{sigd}. Equation~\ref{white_snr} is therefore a valid approximation only when:
\[
\frac{\sigma_0^2}{n} >> \frac{1}{n^2} \sum_{i \neq j} C_{ij} \ .
\]
But in fact, in many cases relevant to planetary transit surveys, this
inequality is not satisfied. Quite the contrary, the covariance term
may even be larger than the white-noise term:
\[
\frac{1}{n^2} \sum_{i \neq j} C_{ij} > \frac{\sigma_0^2}{n} \ ,
\]
so that the full expression in Eq.~\ref{sigcov} should be used to estimate the uncertainty on the transit depth.

Thus Eq.~\ref{white_snr} is no longer valid in the presence of red
noise. The uncertainty on the mean does not decrease as
$n^{1/2}$. Obviously, if neighbouring points are correlated, having
more points during the transit does not increase the detection
statistics as much as if they were uncorrelated.

This can have a drastic effect on the transit detection threshold, as illustrated in~Fig~\ref{wrnoise}. The top curve is drawn with uncorrelated, normally distributed residuals, the middle curve with a red '$1/f$ noise', the bottom curve with a composite noise.  All three curves have identical dispersions. Curves like the bottom panel of Fig.~\ref{wrnoise} will look familiar to observers in the field, while curves like the top panel will look either like the result of an incredibly good night or measurements on rather faint objects for which all other sources of noise are dominated by the statistical photon noise, so that the residuals are uncorrelated and Gaussian.
The middle and bottom curves are much more likely to produce false transit detections (and to hide a real transit signal) than the top curve.

Let us assume a transit duration lasting $50$ times the interval between two points. The dispersion of the mean of $50$ consecutive points in the top curve is $0.14$~($=1/\sqrt{50}$). In the middle curve, this dispersion is $0.40$. Therefore, the middle curve produces a noise for transit detections three times larger than the top curve, even though both curves have the same overall dispersion of the residuals.

\begin{figure}
\resizebox{8cm}{!}{\includegraphics{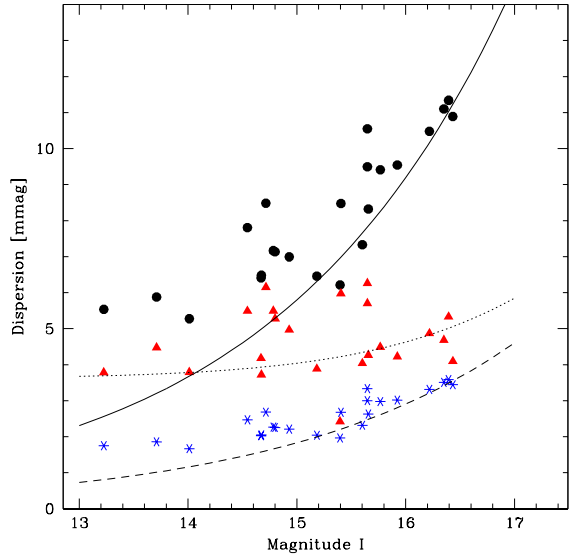}}
\caption{Standard deviation as a function of magnitude for the published candidates in the OGLE survey in the 2001 fields for individual points (filled circles) and for $10$-point averages (triangles). The stars represent the expected position of the $10$-point averages assuming pure white noise ($\sigma/\sqrt{n}$). The solid line is the expected dispersion of individual points due to the white  photon noise, while the dashed line shows the corresponding dispersion for $10$-point averages. For most objects the dispersion of $10$-point averages is much higher than expected for white noise, especially for brighter magnitudes. The dotted line shows the expected dispersion of the $10$-point means according to the discussion in this article, with an amplitude of $\sigma_r=3.6$~mmag for the red noise.  }
\label{msig}
\end{figure}

Figure~\ref{msig} displays, for representative targets of the OGLE
transit survey\footnote{The OGLE survey has been up to now the most
successful survey for transiting planets in terms of detections, $5$
confirmed planets \citep{kon03, bou04, pon04, bou05, kon05} and $177$
published transit candidates. Throughout this article we shall draw
illustrations from the OGLE candidates.}\citep{uda02a,uda02b,uda02c,uda03,uda04}, the behaviour
of the scatter of individual points as a function of magnitude, and
the scatter of $10$-adjacent-point averages ($10$ points span about
$2.5$~hours, a typical transit duration for hot Jupiters), compared
with the expected scatter of $10$-adjacent-point averages in the
presence of uncorrelated white noise. The objects are those of the
2001 season (Galactic bulge). The transit signals were removed
beforehand and only objects with negligible sinusoidal modulations in
the lightcurve were included. Figure~\ref{msig} shows how the
$n^{-1/2}$ decrease of the noise does not apply for most of the survey
objects. The actual dispersion of the mean of $10$ consecutive points
is often much larger than $\sigma_0/\sqrt{10}$, except for the
faintest targets in the survey. For the brightest object, the
$10$-point scatter is even comparable to the $1$-point scatter,
showing that red noise dominates. Since the brightest targets are the
most favourable for the detection of transiting planets,
Fig.~\ref{msig} shows that the white noise assumption is not justified
and an account of covariance must be introduced.

\subsection{Transit signal-to-noise with covariance, $\Sr$}

If we plug Eq.~\ref{sigcov} into Eq.~\ref{white_snr} we get:
\begin{equation}
\Sr \equiv \frac{d}{\sigma_d} = \frac{d}{\sqrt{\frac{\sigma_0^2}{n}+\frac{1}{n^2}   \sum_{i\neq j}  C_{ij}}} \ .
\label{Srdef}
\end{equation}
The $r$ subscript is used to denote the presence of red noise.

In the presence of red noise, $C_{ij}$ for $i \neq j$ is positive
and causes the significance of the transit to increase more slowly
than the familiar $n^{1/2}$ from Eq.~\ref{sigd}. In the limit when 
$\sigma_0$ is small and $n$ is large, the significance may even be totally
dominated by the covariance term.

\subsection{Simple model of the covariance structure}

\label{sgn}

In general, the full covariance matrix of the photometric data is not known, but some reasonable assumptions can be made about it. 
A satisfactory proxy to the effect of the covariance matrix is proposed below, and it is shown to be sufficient to offer a large improvement over the white-noise approximation both to assess the confidence level of a given transit candidate and to estimate the detection threshold of a given transit survey.

For the orbital periods of interest in planetary transit searches, the duration of the transits, noted $l$, is small compared to the period, $P$. Therefore, the data points in the transit will consist of a few stretches of data of duration lower than or equal to the duration of the transit $l$, taken multiples of $P$ days apart.
We make the plausible assumption that the covariance is a monotonically decreasing function of the time difference between two measurements.  Because $l\! <\! <\! P$, the covariance term for points in different transits will be much smaller than for points during the same transit. This implies that to a good approximation the matrix $C$ will  be block-diagonal. If $N_{tr}$ is the number of transits sampled, and $n_k$ the number of points in the $k_{\mathrm th}$ transit, then the significant elements in the covariance matrix will consist of blocks of size $n_k$ by $n_k$: 
\begin{eqnarray*}
 \sum  C_{ij}&=& \sum^{\mathrm{same\ night}} C_{ij} + \underbrace{ \sum^{\mathrm{diff\ nights}} C_{ij}}_{\simeq 0} \nonumber \\
&\simeq &  \sum_{k=1}^{N_{tr}} \sum^{k^{\mathrm{th}} {\mathrm{transit}}} C_{ij} \ .
\end{eqnarray*}
The uncertainty of $d$ in Eq.~\ref{sigcov} will then be:
\begin{equation}
\sigma_d^2 =  \frac{1}{n^2} \sum_{k=1}^{N_{tr}} \sum^{k^{\mathrm{th}} {\mathrm{transit}}} C_{ij} \ .
\label{sumbloc}
\end{equation}

If the sampling interval is constant, then the inner sums in Eq.~\ref{sumbloc} are functions of $n_k$ alone.
Thus, in order to calculate the transit significance $S_r$ in the presence of red noise, it is sufficient to estimate the function $\V$ defined as 
 \[ \V (n) \equiv \frac{1}{n^2} \sum^{n\times n {\mathrm{\ block}}} C_{ij}\]
without the need to fully calculate the individual $C_{ij}$ nor to
make specific assumptions on the dependence of the covariance on the
time separation between two points.
 
For a given lightcurve Eq.~\ref{sumbloc} then becomes:
\begin{equation}
\sigma_d^2=\frac{1}{n^2} \sum_{k=1}^{N_{tr}} \V (n_k) n^2_k \ ,
\label{sig_g}
\end{equation}
and the detection S/N over all individual transits will be
\begin{equation}
S_r^2=  d^2 \frac{n^2}{\sum_{k=1}^{N_{tr}} n_k^2 \V (n_k) } \ .
\label{snr_g}
\end{equation}
The uncertainty on the depth of a single transit will simply be: 
\begin{equation}
\sigma_d= \V^{1/2}(n) \ .
\end{equation}

\subsection{Estimating $\V (n)$ for a given lightcurve}

\label{recip}

We can estimate the function $\V (n)$ for a given transit candidate using the lightcurve data points themselves, $f_i$, by relating $\V (n)$ to the variance of the average of $n$ points in a time interval $l$ outside the transit signal, using the following procedure:

a) remove the points in the transits, as well as any apriori known
systematic effects in the signal;

b) calculate the mean of the flux, $F_j$, over a sliding interval of duration $l$, equal to the duration of the detected transit, recording the number of points $n_j$ in the interval. The interval slides along the whole time series, in steps smaller than the time sampling interval;

c) group the means $F_j$ into bins, according to the number of points in the interval $n_j$;

d) calculate the variance of $F_j$ separately in each bin $n_j$ --  that serves as an estimate of the $\V(n)$ function.

This procedure has the considerable advantage of requiring no external
assumption on the covariance of the residuals. It estimates $\V (n)$
directly from the data itself, preserving the exact covariance
structure of the data as sampled outside the transit. For instance,
photometric trends can be more important at higher airmass, near the
edge of each observing sequence. In that case this effect will be
reflected by an increase in $\V (n)$ for small $n$ (i.e. for cases
when there are only few points in a transit-length interval, which can
only occur at the beginning or end of the observing sequence).

Note that for the simple case of pure white noise (i.e. a diagonal covariance matrix) we expect $\V(n)$ to scale as $1/n$, while for perfectly correlated samples $\V(n)$ will be independent of $n$.

\subsection{Example from the OGLE planetary transit survey}

\label{sogle}

\begin{figure}
\resizebox{8cm}{!}{\includegraphics{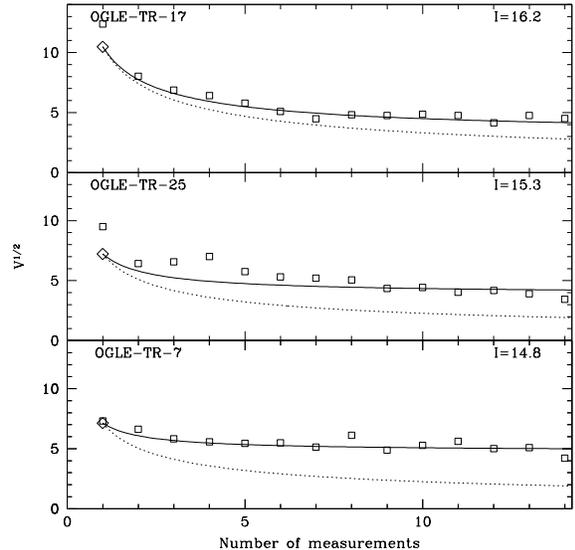}}
\caption{
The $\V(n)$ function for three representative OGLE lightcurves, with brightness increasing from top to bottom. $\V^{1/2}$ (corresponding to the standard deviation of the average of $n$ successive points) is plotted as a function of $n$. In each panel, the empty diamond shows the standard deviation of individual data points in the lightcurve. The dotted line shows a $\sigma/\sqrt{n}$ relation, as expected with uncorrelated noise. The solid line is a fit of $\V(n)=\sigma_w^2/n+\sigma_r^2$ (Eq.~\ref{vnwr}). }
\label{gn}
\end{figure}

We have  applied the above method to some transit candidates published by the OGLE survey. Figure~\ref{gn} displays, by three representative examples, the behaviour of the $\V (n)$ function. We plot the more familiar value $\V^{1/2}$ (corresponding to the standard deviation of the average rather than the variance). 
 On this plot an uncorrelated signal is expected to produce a $n^{-1/2}$ relation, while entirely correlated points will follow a flat relation. Figure~\ref{gn} shows that in the OGLE photometry, sometimes $\V (n)$ evolves quite nearly as  $n^{-1/2}$ as expected for white noise, while in other cases it is almost flat. 
In general, $\V (n)$ decreases less rapidly with $n$ for faint objects than for bright objects. This is obviously related to the fact that as the photon noise becomes larger, it dominates the red noise (the ``systematics''). For the brightest objects, $\V (n)$ is asymptotically constant, which is expected if the points were perfectly correlated. Therefore the noise is completely dominated by low-frequency systematics, which means that gathering more points in a given stretch of time does not add much information.

It is interesting to note that the asymptotic lower limit, towards
which $\V (n)$ seems to converge, is similar for all the objects
regardless of magnitude (see also Fig.~\ref{msig} and
Fig.~\ref{oglesig}).

Note also that $\V (1)$ is not equal to the mean variance of all
points (indicated by empty diamonds in Fig.~\ref{gn}). This is because
$\V (1)$ is estimated only with data points that are isolated in a
duration $l$, i.e. points situated at the beginning or end of a
stretch of measurements. It is quite likely that these points will
tend to be measured at higher airmass values and with atmospheric
conditions varying more rapidly, so that their scatter will be larger
than average.

\subsection{Comparison of the $\Sr$ statistic with the white-noise equivalent}

\label{scomp}

Obviously, one goal in a transit survey is to detect as many
transiting planets as possible. In hypothesis testing jargon, we
basically wish to reduce the probability of type-I error (false
negative). However, this means reducing the detection threshold, which
in turn implies increasing the type-II error (false positive). The
number of false positives is the practical bottleneck, since it bears
an immediate implication on the amount of follow-up observations
needed. We therefore have to fix the threshold according to the amount
of false positives we are willing to tolerate, taking into account the
size of the survey.

\begin{figure}
\resizebox{8cm}{!}{\includegraphics{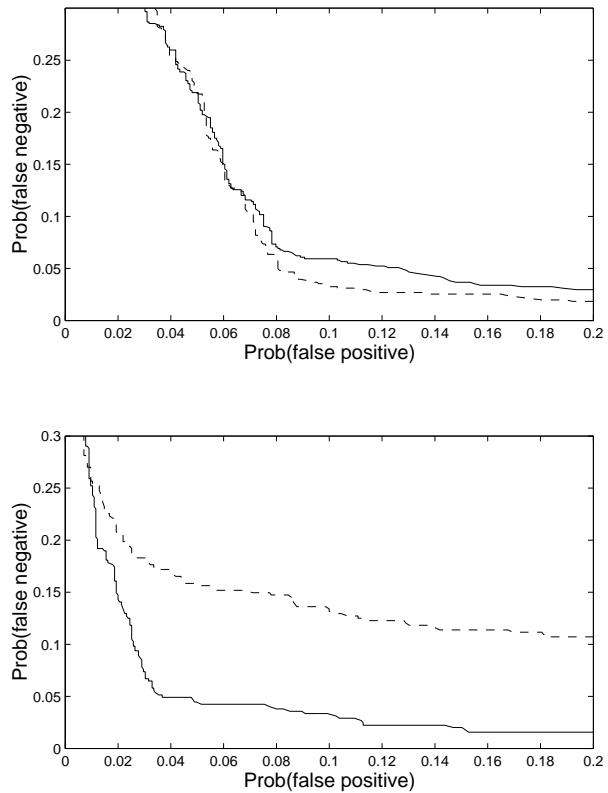}}
\caption{Type-I (false negative) and type-II (false positive) errors
for the two simulated datasets with pure white noise (upper panel) and
extremely red noise (lower panel). The curves are produced by varying
the detection threshold, which controls the two types of error rates. The
dashed curve represents the performance of $S_d$ and the solid curve
that of $\Sr$.  }
\label{errtypes}
\end{figure}

In order to compare the performance of the $\Sr$ statistic to that of
white-noise statistic $S_d$, we produced two synthetic datasets: each
dataset consisted of $1000$ lightcurves with no transits, and $1000$
lightcurves with transits of varying periods, phases, depths, and
noise levels. The simulated observations lasted for $50$ nights, with
a sampling interval of $15$ minutes. The two datasets differed by the
way the noise was generated -- the first dataset had a pure white
noise, while the second had a strongly covariant noise with an
exponentially decaying correlation -- $\rho(\Delta t) =
e^{-\frac{\Delta t}{\tau}}$, with a time constant of
$\tau=13$~minutes.  At the starts and ends of nights the correlation
was artificially increased to simulate high airmass effects. No
inter-night correlation was introduced.  We ran the BLS detection
algorithm \citep{kov02} on the two datasets and computed $S_d$ and
$\Sr$ for all the lightcurves. A lightcurve for which the true period
was found by BLS (to reasonable accuracy) was tagged as a detection.
By varying the detection threshold we change the error rates of both
types.  Figure \ref{errtypes} shows the behaviour of the two error
types for the two datasets. The dashed curve represent the performance
of $S_d$ while $\Sr$ is represented by the solid curve. In the case of
pure white noise (upper panel) the difference between $S_d$ and $\Sr$
is marginal. However, in the case of the red-noise dataset (lower
panel), at each level of type-II error (false positive) rate we can
reach a considerably lower type-I error (false negative) rate using
$\Sr$.

\subsection{Single-parameter description of the covariance with $\sigma_r$}

\label{Ssigr}

\begin{figure}
\resizebox{8cm}{!}{\includegraphics{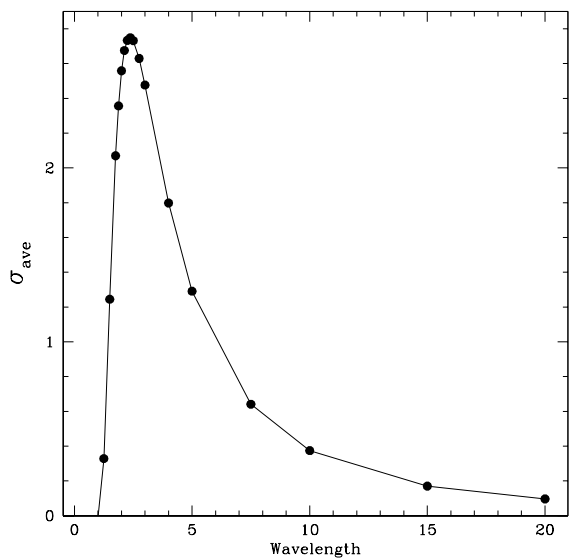}}
\caption{Standard deviation of the average over a duration $l$ for sine signals of unit dispersion and different wavelengths. The wavelength is expressed in units of $l$. }
\label{sines}
\end{figure}

To model the behaviour of $\V (n)$ in the presence of red noise, let us represent the red noise as a Fourier sum of sine curves. For a sine signal, the dispersion of the mean of a set of points over a duration $l$ is a very strong function of the wavelength of the sine signal. Figure~\ref{sines} gives this dispersion as a function of the wavelength. This dispersion is the amplitude of the systematics that will remain in the signal regardless of the number of points measured during an interval of duration $l$. It is seen to peak at wavelengths slightly above twice the duration of the transit. Schematically, the correlated noise can be thought of as consisting of three components: short frequencies that will tend to average out over the duration of the transit, long frequencies that will not vary between the transit and the neighbouring measurements, and frequencies around $(2 l)^{-1}$ that will introduce strong transit-like residuals. 

Let us assume then that the noise can be separated into a purely white component, noted $\sigma_w$, and a purely red component, noted $\sigma_r$  (where $w$ and $r$ stand for ``white'' and ``red''), expressing the power of the correlated noise at frequencies near $(2 l)^{-1}$.

$\sigma_w$ results from several independent white noise components
such as photon noise, sky noise and scintillation. In general it is a
function of the target magnitude. $\sigma_r$ results from systematics
(correlated noise). Both parameters $\sigma_w$ and $\sigma_r$ can be
derived from a given lightcurve from the behaviour of $\V(n)$. For
purely white noise, the variance scales as $\V(n)=\sigma^2_w / n$,
while for purely red noise it scales as $\V(n)=\sigma^2_r$ (see
Section~\ref{recip} and Fig~\ref{gn}). In intermediate cases, we can
model it as $\V(n)=\sigma_w^2 /n+\sigma_r^2$, so that it is near
$\sigma^2_w /n$ for individual points and approaches $\sigma^2_r$ as
$n$ grows larger.

\begin{figure}
\resizebox{8cm}{!}{\includegraphics{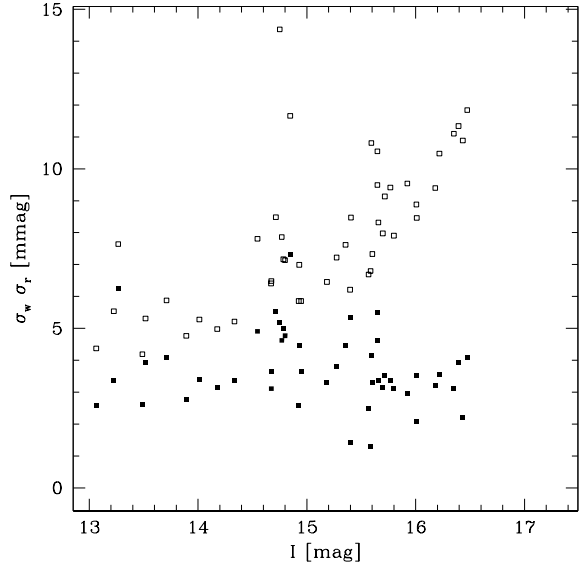}}
\caption{$\sigma_w$ (white symbols) and $\sigma_r$ (black symbols) for OGLE candidates from the 2001/2 fields. The amplitude of $\sigma_r$ is found to be of a few millimagnitudes in almost all cases. In fact, given the uncertainty in the determination of $\sigma_r$ itself (involving a quadratic subtraction), values of $\sigma_r$ for the different targets are remarkably similar. It therefore seems that to a fair approximation, all targets in the survey can be thought of as affected by a similar red noise component, largely independent on magnitude. The mean value is $\sigma_r=3.6$~mmag for the 2001/2002 targets (Galactic bulge) and $\sigma_r=3.1$~mmag for the 2002/2003 targets (Carina).}
\label{oglesig}
\end{figure}

We have calculated the values of $\sigma_w$ and $\sigma_r$ for published candidates in the case of the OGLE survey. The results are displayed in Figure~\ref{oglesig}. The amplitude of the white noise has a familiar dependence on magnitude, with photon noise dominating except at the bright end. The red noise shows no dependence on magnitude. Its mean is $3.6$~mmag for the 2001/2 season and $3.1$~mmag for the 2002/3 season. The difference is probably due to a less crowded field and improvements in the reduction procedure for the second season.

Using published information, we infer that values
near $3$~mmag for $\sigma_r$ are typical of some other surveys as well. 
For instance, in a transit survey centered on NGC 2301 with a
$2.2$m telescope, \citet{ton05} find a red noise of $\sim 3$~mmag (see
their figures 4 and 5)\footnote{Note that intrinsic variability is
also included in the red noise in addition to systematics in the
photometry. \citet{ton05} find a very high occurence of variability at
the millimag level. These have the same effect as systematics on the
detection of transits.}. Assuming that red noise dominates for the
brightest targets, similar values of $\sigma_r$ are also inferred from
the magnitude-dispersion plots and transit candidate depths for
several other surveys cited in the Introduction.

\subsection{Parameter dependence of the $\Sr$ statistic}

\label{ssds}

In order to estimate $\V (n)$ and $\Sr$ 
when the actual lightcurves are not available, we use the expression
of $\V (n)$ from the previous section:
\begin{equation}
\V(n)= \sigma^2_w / n + \sigma^2_r \ .
\label{vnwr}
\end{equation}

We can then write:

\begin{equation}
\sigma_d^2= \frac{1}{n^2} \sum_{k=1}^{N_{tr}} n_k^2 \left( \frac{\sigma_w^2}{n_k} + \sigma_r^2 \right) \ ,
\label{sig_wr}
\end{equation}
and express $\Sr$ as:
\begin{equation}
\Sr^2 = \frac{d^2 n^2}{ \sum_{k=1}^{N_{tr}} n_k^2 \left( \frac{\sigma_w^2}{n_k} + \sigma_r^2\right)} \ ,
\label{snr_wr}
\end{equation}
where $N_{tr}$ is the number of transits sampled, $n_k$ is the number of data points in the $k^{\mathrm th}$ transit.

To relate Eq.~\ref{snr_wr} to the transit detection threshold, we express it in terms of physical parameters. For simplicity we assume here a homogeneous distribution of data points in phase, so that $n_k \simeq n/N_{tr}$ for all transits sampled. Then Eq.~\ref{snr_wr}  becomes:
\begin{equation}
\Sr^2 = \frac{d^2}{\frac{\sigma_w^2}{n} + \frac{\sigma^2_r}{N_{tr}}} \ .
\label{snr_rh}
\end{equation}

If $N$ is the total number of data points per star, and $\delta_t$ the typical time interval between two measurements, then the number of points during the transit is $n=\beta N R/\pi a$ (with $a= c P^{2/3} M^{1/3}$), and the mean number of transits sampled is $N_{tr}=N \delta_t /P$,
 where $a$ is the orbital semi-major axis (assuming a circular orbit), $P$ the orbital period, $M$ and $R$ the mass and radius of the primary, $c$ is a constant equal to $1$ in Sun-Earth units (\Msol, years, AU), and $\beta$ is a correction factor due to the latitude of the transit ($\beta=1$ for a central transit).

Then  Eq.~\ref{sig_wr} becomes:
\[
\sigma_d^2 = c \pi \frac{\sigma_w^2 P^{2/3} M^{1/3}}{\beta N R} + \frac{\sigma_r^2 P }{N \delta_t} \ ,
\]
and Eq.~\ref{snr_wr}:
\begin{eqnarray}
\Sr^2 &=& d^2/\sigma_d^2 = \alpha (r/R)^4 /\sigma_d^2 = \nonumber \\
 & &  \alpha (r/R)^4 \left[ c \pi \frac{\sigma_w^2 P^{2/3} M^{1/3}}{\beta N R} + \frac{\sigma_r^2 P }{N \delta_t}\right]^{-1}
\label{snr_phys}
\end{eqnarray}
is the detection significance in the presence of systematics, expressed as a function of the characteristics of the planetary system. $r$ is the radius of the planet and $\alpha$ a parameter dependent on limb darkening, accounting for the fact that actual transit signals are a bit deeper than $(r/R)^2$ because of limb darkening: $d=\alpha (r/R)^2$.

The limit of this expression in the uncorrelated noise regime ($\sigma_w/\sqrt{n}>>\sigma_r$) is 
\begin{equation}
\Sr = \left( \frac{\beta \alpha}{\pi c}\right) ^{1/2} (r/R)^2 N^{1/2} R^{1/2} P^{-1/3} M^{-1/6} \sigma_w^{-1} \ ,
\label{snr_w}
\end{equation}
which corresponds to the white-noise expression Eq.~\ref{white_snr}, while the limit in the correlated regime is
\begin{equation}
\Sr=\alpha^{1/2}  (r/R)^2 N^{1/2} \delta_t^{1/2} P^{-1/2} \sigma_r^{-1} \ .
\label{snr_r}
\end{equation}

Numerically, expression~\ref{snr_r} is very different from expression~\ref{snr_w}. For typical parameters of transit surveys (e.g. $R\! =\! 1$~\Rsol, $M\! =\! 1$~\Msol, $N\! =\! 1000$, $\delta_t\! =\! 15$~min, $\alpha\! \simeq \! 1$, $P\! =\! 3.5$~days, $r\! =\! 0.1$~\Rsol, $\sigma_w \! =\! 0.003$, $\sigma_r\! =\! 0.003$),  $\Sr$ is about $3$ times larger than predicted by white noise alone. This has profound implications for the yield prediction and interpretation of transit surveys.

Eq.~\ref{snr_rh} can be expressed in terms of its white-noise equivalent. A bit of algebra gives:
\[
\Sr= S_w [1+(\sigma_r/\sigma_0)^2 n_k]^{-1/2} \ ,
\]
where $S_w \equiv d/\sigma_0 \sqrt{n}$ is the white-noise significance. When red noise dominates, this reduces to 
\[
\Sr \simeq S_w n_k^{-1/2} \ ,
\]
i.e. the significance is reduced by the square root of the typical
number of points in an individual transit. For the parameters above,
$n_k \sim 10$. Hence the result that the detection threshold is
about three times higher for bright targets in the OGLE survey than indicated by the
white-noise approach (see Section~\ref{appogle}). Other surveys
typically have a smaller $\delta_t$, so that the reduction factor is
even larger.

\subsection{Generalisations}
\label{app}

In many cases the uncertainties of the individual data points are not
constant. An additional assumption is then needed about the relation
between the white noise and the red noise. A simple way to generalise
our method is to assume that the red and white noise are
proportional. This is reasonable, for instance, if higher
uncertainties are due to a higher airmass or poor weather conditions,
which are expected to increase both the white and red noise. In that
case, the equivalent of Eq.~\ref{snr_g} can be computed from the
variance of the weighted mean of the flux in a transit-length
interval. However, there may be situations when the red noise is not
proportional to the white noise. For instance, if the exposure time 
changes by a significant factor, the photon noise will change, but the
photometric systematics may remain unaffected. In that case, a
possible solution is to parametrize $\V$ as a combination of red and
white noise as in Section~\ref{Ssigr}, and to determine $\sigma_r$ on
the assumption that $\sigma_w$ is proportional to the photon noise. In
most cases, though, it may be sufficient to assume a mean constant
uncertainty to compute $\V$ even if the real uncertainties somewhat vary.



Equation~\ref{snr_g} can also be generalised to account for unequal
depths in the different transit sampled. Instead of using the same $d$
for all transits in Eq.~\ref{Srdef} and \ref{snr_g}, the S/N can be
summed over all the transits using the individual $d_k$, the mean
depths of the data in individual transits.

Some transit detection algorithms look for single transits by searching for sharp flux changes rather than by fitting a step function to the phase-folded signal. Such methods can perform better when photometric trends within and between nights are strong. In that case, $\V (n)$ should be computed by using the variance of the mean flux difference between two neighbouring stretches of data (as used by the detection algorithm) rather that the variance of the mean flux during the transit. This may also be a suitable approximation when the detection is performed by visually inspecting the lightcurves.

\section{Applications}

\label{ssurveys}

\subsection{Selecting transit candidates in a survey}

\label{sthres}

The $\Sr$ statistic is a significance indicator for potential transit signals that, as discussed in Section~\ref{scomp}, is superior to the usual S/N statistics in the presence of covariant noise. It can therefore be used to estimate the significance of a transit candidate, i.e. the estimation of the probability that the detected signal is due to random fluctuations rather than real transits.

Existing surveys have shown that transiting planets produce signals
that are near the detection threshold, and that it is important to
include candidates that are as near as possible to the spurious
detections. However, the presence of red noise makes this
difficult. Most surveys have tackled this problem by staying well
clear of the minimum threshold, and retaining only clear signals with
high signal-to-noise for subsequent follow-up and confirmation. But
this comes at the price of missing potentially interesting
candidates. The use of the $S_r$ statistic to evaluate the
significance of the transit detections should allow a lower detection
threshold and better identification of spurious signals. It is also
useful for ranking the candidates before the demanding follow-up
efforts.

The OGLE survey has been striving to set the detection threshold as
low as possible, even at the risk of including a few spurious
detections. Here we apply our formalism to the OGLE candidates and
show that it provides a more reliable definition of the effective
detection threshold and a better discrimination of false positives.

\label{appogle}

\begin{figure}
\resizebox{9cm}{!}{\includegraphics{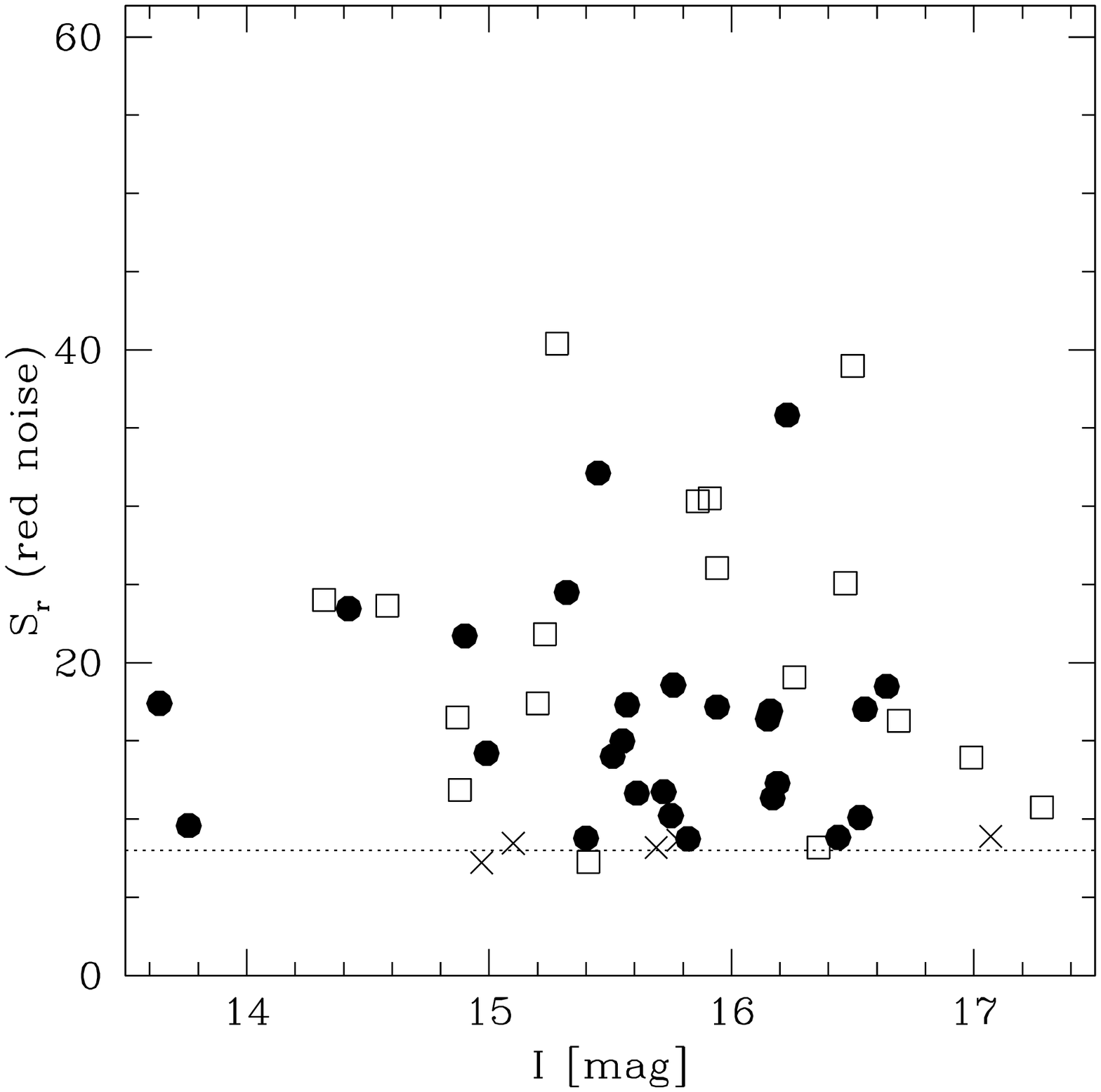}\includegraphics{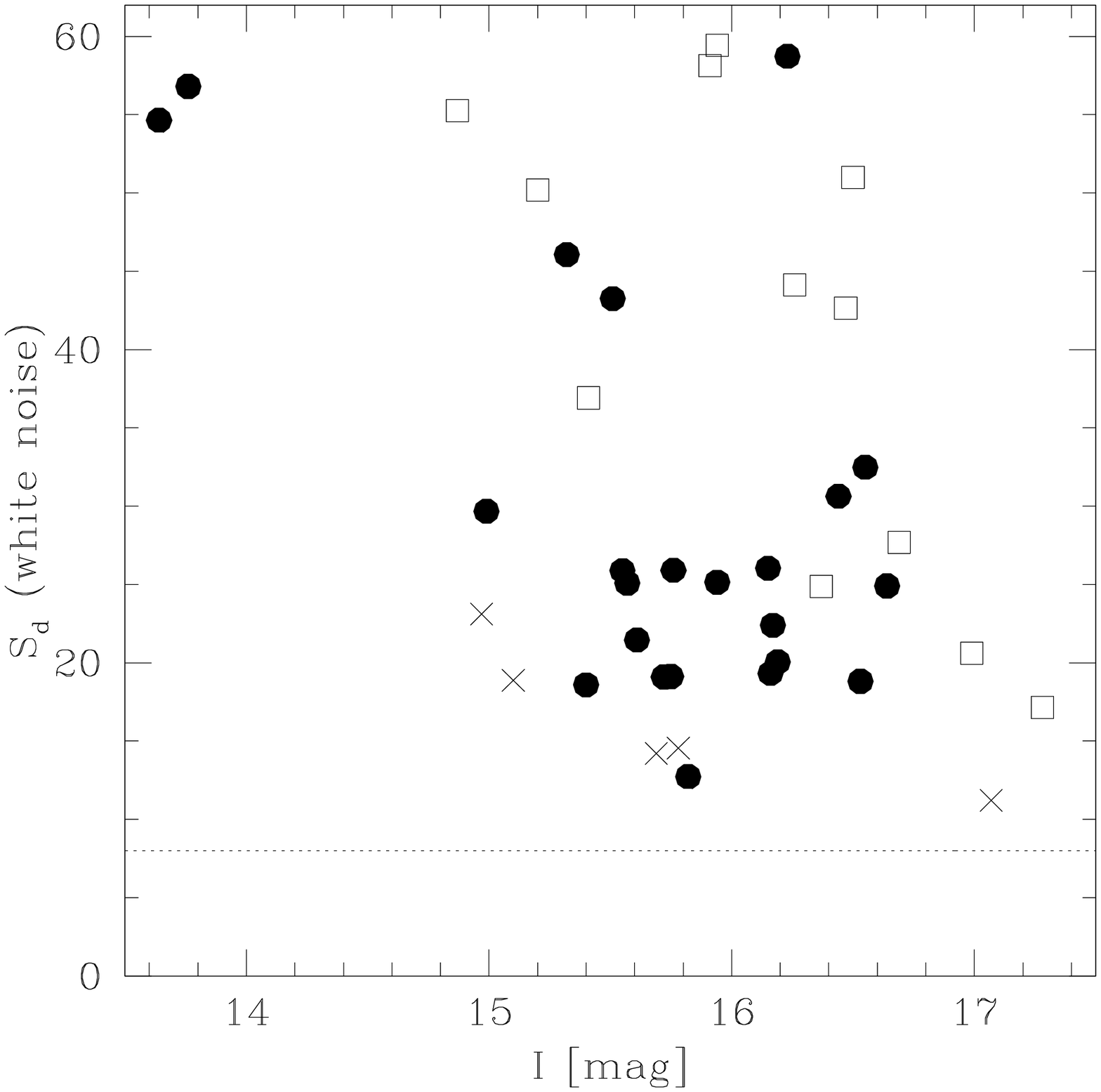}}
\caption{ {\bf Left--} $S_r$ versus magnitude for the OGLE candidates of the 2002/3 season. Filled circles indicate candidates confirmed by the radial velocity or photometric follow-up as bona fide transits or eclipses, crosses stand for the possible false positives according to the spectroscopic follow-up, and squares mark the candidates showing obvious sinusoidal modulations in the lightcurve. {\bf Right--} same axes and symbols, with transit signal-to-noise calculated assuming uncorrelated noise. The dotted line shows a threshold signal-to-noise value of $8.$ }
\label{oglesd}
\end{figure}

Using the $\V (n)$ function calculated for each OGLE transit candidate, we computed the $\Sr$ indicator with Eq.~\ref{snr_g} for the published OGLE transit candidates. 
The left panel of Fig.~\ref{oglesd}  displays the results for the 2002/3 observing season (OGLE-TR-60 to OGLE-TR-132), for which the radial velocity follow-up is essentially complete \citep{pon05}. For comparison, the right panel shows the white-noise detection signal-to-noise $S_d$. Crosses indicate suspected false positives according to the spectroscopic follow-up, and filled circles the confirmed eclipsing binaries and transiting planets.
The apparent threshold  in the $\Sr$ plot, around $\Sr=8$, constitutes  a much sharper detection criterion than $S_d$. Of the nine objects with $\Sr<9$ among the candidates of the first two seasons (OGLE-TR-48, TR-55, TR-58, TR-89, TR-118, TR-124, TR-125, TR-127 and TR-131), seven were shown by the spectroscopic follow-up and further photometric measurements to be likely false positives \citep[][A. Udalski, priv. comm.]{bou05,pon05}. Only OGLE-TR-55 and OGLE-TR-125 were confirmed as bona fide eclipsing binaries. 

The white-noise $S_d$ offers a much poorer separation of the false
positives, especially for the brighter targets.  As expected, the $\Sr$
statistic eliminates the strong dependence of the threshold on
magnitude, since systematics become the dominant source of noise for
the brighter targets. 
The bright candidates
most likely to be false positives according to the radial velocity
follow-up get a much lower ranking, closer to the detection threshold
near $\Sr=8$. At the bright end of the magnitude range, the
significance decreases by a factor $\sim 3$ for OGLE targets compared
to the white-noise analysis, which is approximately the square root of
the average number of data points in an individual transit,
illustrating the transition from the uncorrelated regime to the
heavily correlated regime.

These results confirm the value of $\Sr$ as a robust confidence
indicator, superseding white-noise signal-to-noise indicators.  It
also shows that thresholding $\Sr$ is a very good model for the
combination of white signal-to-noise, spectral 'SDE' index of the BLS
and by-eye selection performed by the OGLE team.

For the 2004 season \citep[OGLE-TR-138 to TR-177,][]{uda04}, the OGLE
team lowered the acceptance threshold for the published transiting
candidates, in an attempt to detect more transiting planets. The
spectroscopic follow-up of these candidates has not been completed to
date. It is therefore useful to apply our approach to these new
candidates.We find that OGLE-TR-143, TR-150, TR-152, TR-160, TR-161,
TR-162, TR-166, TR-168, TR-172, TR-173, and TR-174 have $\Sr<8$ and
are therefore highly likely to be false positives.  As a consequence,
the absence of a spectroscopic eclipsing binary signature for these
objects will not be a compelling indication that they host a
transiting planet.

\begin{figure}
\resizebox{8cm}{!}{\includegraphics{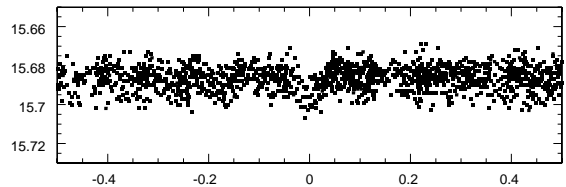}}
\caption{Phase-folded lightcurve of OGLE-TR-161 as an example of detection with low $S_r$ ($S_r=5.5$). Note the structure outside the candidate transit signal.}
\label{lowsd}
\end{figure}
  
By examining published transit candidates and inferring plausible values
of $\sigma_r$ from published information, we have found clear
indications that the $\Sr$ statistic also provides useful results when
applied to other surveys. For instance, the TrES-1 transiting planet
detected by the TrES network has $S_d \sim 35$ and $\Sr \sim 12$. Its
high white-noise significance would make it difficult to understand
why such a deep transit was the only planet detected by the TrES
survey -- given that shallower planetary transits are expected to be
much more abundant. However, the relatively low $\Sr$ indicates that
it may be situated near the actual detection threshold.

\subsection{Improved error estimates}

The expression of Eq.~\ref{sig_g} for $\sigma_d$ also gives an estimate of the uncertainty on the depth of the transit, taking into account the correlation in the noise. This expression can replace uncertainty estimates based on white-noise $\chi^2$ statistics that are likely to yield underestimated error intervals. In turn, the  $\sigma_d$ can be used to derive the uncertainties on the stellar and planetary parameters.

This method is in essence similar to the ``residual permutation''
method that we used in~\citet{bou05} and \citet{pon05}, earlier
suggested in~\citet{jen02} under the name ``segmented bootstrap''.

This method can also be extended to compute realistic uncertainties on other parameters than transit depth, such as the duration, orbital inclination or central epoch. This can be done by using the $\V $ function to estimate the expected variance between the observed data and the actual value, using only the points that are contributing to the determination of the parameter in question (for instance, for the transit epoch, the points during the transit ingress and egress). The value of the parameter can be moved away from the best-fitting value until the sum-of-squares is equal to the variance expected from Eq.~\ref{sig_g}.

We have used this method on the multi-colour, multi-site transit photometry of HD189733 to show that the white-noise uncertainties were underestimated and that correlated noise could account for unexplained features in the results \citep{bak06}.

\subsection{Evaluating the potential of planetary transit surveys}

In order to evaluate the potential of a given planetary transit
survey in terms of planet detection (e.g. at the planning stage), a
model of the transit detection threshold is required. In previous
sections we showed why a threshold in the $\Sr$ statistic can provide
a significant improvement over the white-noise statistic generally
used for this purpose.

\subsubsection{Numerical value of the $\Sr$ threshold}

\begin{figure}
\resizebox{8cm}{!}{\includegraphics{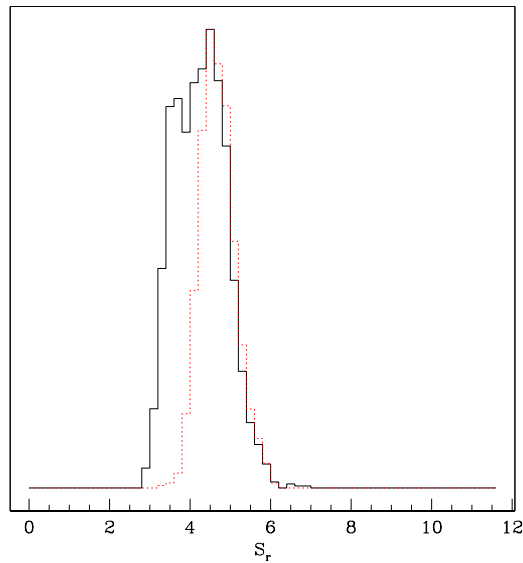}}
\caption{Histogram of $\Sr$ for $1000$ simulated lightcurves without transit signals, with a time sampling of $15$ minutes over $50$ nights. Solid line: red+white noise, dotted line: white noise.}
\label{simshay}
\end{figure}

The numerical value of the detection threshold in terms of $\Sr$ will
depend on the time sampling of the lightcurve and on the desired rate
of false alarms considered acceptable. In the OGLE survey in Carina
for instance, we find that false alarms become predominant over
positive detections for $\Sr<9$ (see Section~\ref{appogle}). The
lowest confirmed planetary transit has $\Sr=11.6$, and the lowest
confirmed eclipsing binary has $\Sr=8.8$. In our simulations for $50$
consecutive nights with white, red and entirely correlated noise, the
threshold $\Sr$ values for false positives are very rarely larger than
$7$ (see Fig.~\ref{simshay}), although correlated lightcurves do produce
false positives up to $\Sr=12$ at rates of a few per thousands. The
white-noise values are consistent with threshold values estimated by
\citet{jen02} between $5.5$ and $7.5$ for surveys spanning $1$ week to
$4$ years, using the white-noise analogous to $\Sr$.  Published
information about transit candidates from other surveys than OGLE also
imply similar values of the threshold.

In summary, the effective detection threshold in actual cases will depend on the time sampling, desired false alarm rate and the detection algorithm used. It will typically be in the $7$--$9$ range.

\subsubsection{Behaviour of an $\Sr$ threshold}

An $\Sr$ threshold has a different behaviour than the white-noise $S_d$ generally used. $\Sr$ is not proportional to the dispersion of individual points $\sigma_0$, it ``saturates'' before $\sigma_0$ leaves the photon-noise regime towards the systematic-dominated regime. For values of $\sigma_r$ in the range of a few millimag, this implies that most planetary transits will stay below the detection threshold regardless of the magnitude, with important implications for the detection potential of transit surveys. 
 
\begin{figure}
\resizebox{8cm}{!}{\includegraphics{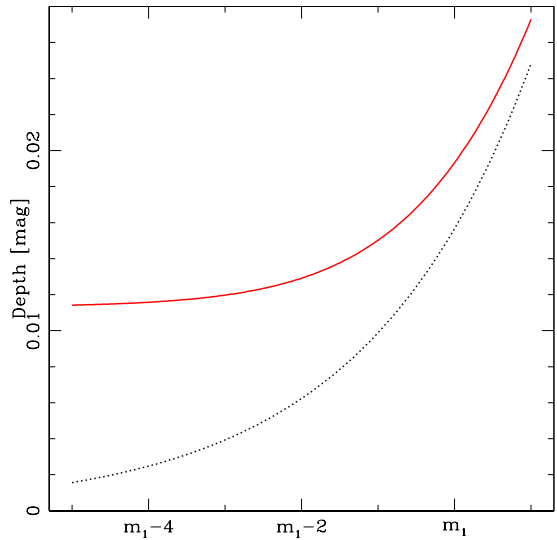}}
\caption{
Detection threshold in terms of transit depth as a function of
magnitude, with a white noise dominated by photon-noise and
scintillation and a red noise characterised by $\sigma_r=3$~mmag
(solid line) and white noise only (dotted line) for a typical planetary
transit signal with $P=3.5$~days. $m_1$ is the magnitude for which the photon noise is 0.015 mag.}
\label{magd}
\end{figure}

\begin{figure}
\resizebox{8cm}{!}{\includegraphics[angle=-90]{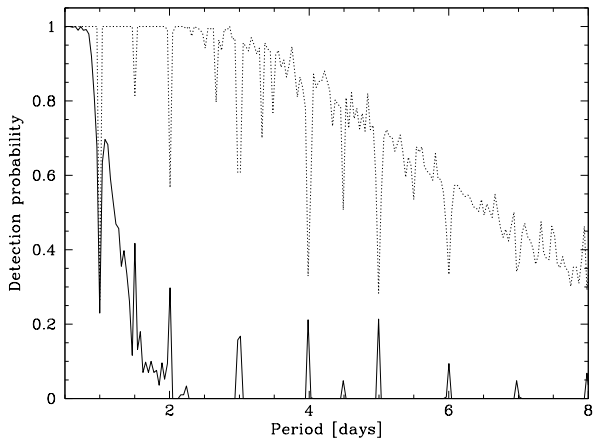}}
\caption{Detection probability as a function of period, with $\sigma_r=3.1$~mmag (solid curves) and $\sigma_r=0$ (dotted curves) for a primary radius of $1.0$~\Rsol\ and a survey durations of $\sim$ $50$ nights. The secondary radius is fixed to $0.11$~\Rsol, $\sigma_w$ is set to 3 mmag. The detection criterion is $\Sr>8$ and the presence of data in at least three different transits.}
\label{probdec}
\end{figure}

 Figure~\ref{magd} plots the detection threshold in terms of transit depth (from Eq.~\ref{snr_phys}--\ref{snr_r}) with a photon noise of $0.01$~mag at the $m_1$ magnitude, a scintillation noise of $2$~mmag and no sky background noise.
The dotted line shows the equivalent relation if the noise is assumed
to be white. Figure~\ref{magd} shows that taking red noise into
account makes an enormous difference in the detection threshold. The
difference is largest for the brighter targets, which are also the
most important targets for planet detection. Since most hot Jupiters
produce transits of depth $1\%$ or shallower, and have periods of
$3$--$4$~days or higher, one conclusion of Figure~\ref{magd} is that
red noise will make transit surveys much less efficient in terms of
planetary detection than would be indicated by white noise
estimates. The loss of efficiency will be even higher than suggested
by a simple comparison of the detection threshold, since the presence
of red noise causes the detectability zone to stay constantly above
the typical depth of hot Jupiter transits even for the brightest stars
in the survey. Therefore, detections will be confined to peculiar
cases such as exceptionally deep transits or favourable periods (see Fig.~\ref{probdec}) --
precisely as observed in the characteristics of the seven transiting
planets detected to date by transit surveys.

The introduction of red noise in Eq.~\ref{snr_phys} has another
interesting consequence: it introduces a somewhat steeper dependence
of detectability on period than in the case of white noise. The reason
is that for short periods, an equivalent number of data points during
the transit will be distributed over a larger number of transits,
mitigating the effect of the covariance. Figure~\ref{probdec} shows
the results of Monte Carlo simulations computing the proportion of
detected transits as a function of periods, with the setup of the OGLE
survey ($1100$ points over about $60$ nights), at the bright limit
($\sigma_0\simeq \sigma_r$), with $\sigma_r=3.1$~mmag and a threshold
for detection $S_T=8$. The presence of red noise introduces a sharp
decline of the detectability above a rather short period. For
instance, for a Solar-type primary, the detectability is negligible
above $P=2$~days. This contributes to explain the apparent
inconsistencies of the results of the OGLE survey (3 Very Hot Jupiters
with P=$1$--$2$~days, for $2$ hot Jupiters) compared to radial
velocity surveys (more than two dozens hot Jupiters and no $P<2$ days
planet). The period dependence of detectability in the OGLE survey is
even steeper than modelled by \citet{gau05} with a white-noise
approximation (see next section for quantitative estimates).

In our simulations as well as in the OGLE data, we also found a
residual dependence of the effective $\Sr$ detection threshold on
period. This is due to the fact that detection algorithms have more
difficulty finding low signal-to-noise transits at longer periods, at
a given $S_r$ level, because with less transits sampled there are
more possible ``foldings'' of real transits with systematic
fluctuations to produce false detections.

\subsubsection{Detection potential of transit surveys with a $\Sr$ threshold}

The potential of a planetary transit survey will depend not only on
the detection threshold but also on the observed target
populations. These will be specific to each case and require specific
simulations. Without going into the details of any specific survey, it
is interesting here to examine the effect of the amplitude of the red
noise, expressed by $\sigma_r$, on the potential of a transit survey
with typical parameters. We have run a simulation for a ``typical''
survey consisting of $50$ nights of observations of $10000$ targets
with photon noise lower than $0.01$~mag. We assumed that $1\%$ of the
targets have a hot Jupiter, with a log-flat distribution of periods between 3 and 10 days,
 that half of the targets are binary or blended, and used star/planet radius ratios drawn from a log-normal
distribution centered on $-1.15$ with dispersion $0.18$, and a sampling interval of $5$ minutes.

\begin{description}

\item{\em Effect of systematics on the total yield}\\
 Figure~\ref{potential} shows the results in terms of expected number of transiting hot Jupiter detections as a function of the value of the $\sigma_r$ parameter, assumed to be identical for all targets. It illustrates the drastic effect of the amplitude of $\sigma_r$ on the potential of ground-based surveys. For values of $2$--$3$~millimag, the expected yield is reduced by a factor $2$ to $4$ compared to the white-noise expression. Photometry with red noise above a few millimag cannot be expected to provide a significant contribution to the detection of transiting planets.

\begin{figure}
\resizebox{8cm}{!}{\includegraphics{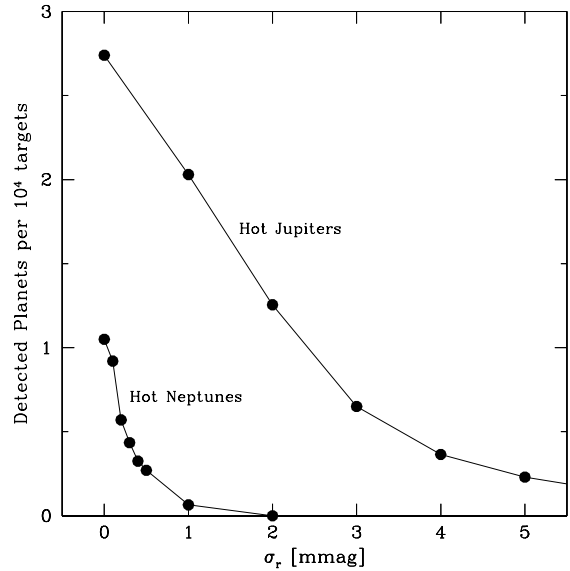}}
\caption{Number of transiting hot Jupiters (top curve) and hot Neptunes (bottom curve) found in our ``typical'' ground-based transit survey simulation, as a function of the $\sigma_r$ red-noise parameter. Hot Neptunes are assumed here to be ten times more abundant than hot Jupiters, and to have radii around $0.3 R_J$.}
\label{potential}
\end{figure}

\vspace{-1mm} \ \\

\item{\em Hot Jupiters vs. Very Hot Jupiters}\\
We examined the effect of changing the period distribution of the
transiting planets. We considered the case of ``very hot Jupiters'',
with periods between $1.2$ and $2$~days. Assuming white noise, such
planets are more easily detected, because their transits occur more
frequently and also because the geometric probability of a transit
configuration is higher \citep[see][]{gau05}. We find that with
$\sigma_r$=3 mmag, the detectability of very hot Jupiters compared to
$3$--$4$~days hot Jupiters is further increased by a factor $\sim 2$, and
by a factor $\sim 3$ with $\sigma_r=5$ mmag. This helps to
explain the marginal mismatch between the period distributions of hot
Jupiters detected with radial velocities and with photometric transit
surveys.

\vspace{-1mm} \ \\

\begin{figure}
\resizebox{8cm}{!}{\includegraphics{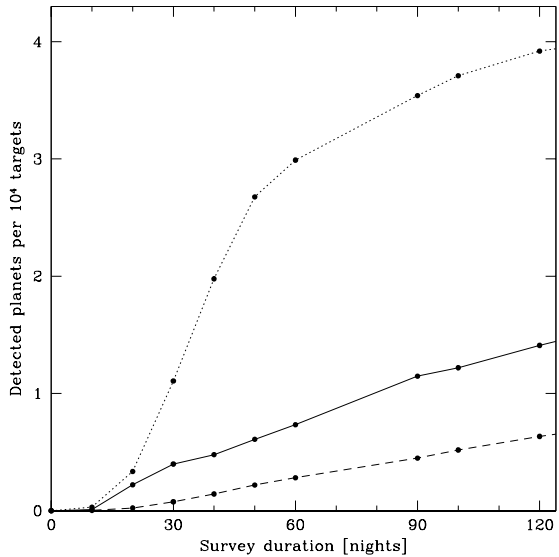}}
\caption{
Expected number of transiting hot Jupiter detections in our "typical" survey as a function of the total number of nights of observation, for purely white noise (dotted line), and two different levels of red noise: $\sigma_r=3$ mmag (solid line) and $\sigma_r=5$ mmag (dashed line).}
\label{duration}
\end{figure}

\item{\em Total survey duration}\\
We also considered the effect of modifying the total duration of the
transit survey. The simulations indicate that the presence of red
noise lengthens the duration of the survey necessary before the number
of detections starts to saturate (see Fig.~\ref{duration}). 
With purely white noise, the number
of detections starts to saturate after about $50$ nights, but with
$\sigma_r$=3-5 mmag it keeps rising almost linearly beyond $100$
nights. This indicates that correlated noise moves the optimal
length of a transit survey towards longer campaigns on the same target
fields. Whereas white-noise estimates may indicate that $30-50$ nights of
measurements on a field before moving to another one is the optimal
strategy, taking red noise into account moves the optimum towards
larger campaigns. The basic reason behind this result is that, because
of the trends in the photometry, a higher number of individual
transits may be required to achieve the necessary signal-to-noise for
a secure detection. For instance, for the five detected OGLE
transiting planets the number of individual transits sampled is $7$, $12$,
$9$, $10$ and $11$ -- whereas simulations indicate that most detections should
be based on $3$--$5$ transits. The expected yield of a survey therefore
keeps rising until the most common hot Jupiters, with periods of $3$--$4$
days, can be sampled enough times.

\vspace{-1mm} \ \\

\item{\em Hot Neptunes}\\
Several claims have been made about the capacity of ground-based surveys with big telescopes to detect ``Hot Neptunes'' i.e. close-in planets with sizes around $0.3$~$R_J$, all based on the white-noise approximation \citep[e.g. ][]{har05,gil05}. To examine what value of $\sigma_r$ such detections would require, we repeated the same simulation with planets of Neptune size ($\sim 0.3$~$R_J$). We assumed that such planets are ten times more abundant than hot Jupiters. Figure~\ref{potential} shows that extremely small values of $\sigma_r$ would be required to attain a significant efficiency in the detection of Hot Neptunes, $\sigma_r<0.3$ mmag. This would be exceedingly difficult to achieve in wide-field photometry over long time scales. Therefore the photometric detection of transiting Hot Neptunes from the ground requires an unlikely leap in the treatment of systematic effects in the photometry. 

\end{description}

\section{Summary and implications}

This study shows that in ground-based surveys for planetary transits,
the systematics in the photometry have a very important effect on the
capacity to detect planetary transits, and that estimates of
detection significance and detection threshold based on the assumption
of white Gaussian noise are not appropriate. We propose a simple formalism
to assess the significance level of detected transits, and
to predict the detection threshold of an observing campaign.

We find that the uncertainty on the depth $d$ of a transit candidate is
\begin{equation}
\sigma_d^2= \frac{1}{n^2} \sum_{k=1}^{N_{tr}} n_k^2 \V (n_k) \ ,
\end{equation}
where $N_{tr}$ is the number of transits sampled, $n_k$ the number of data points in the $k^{\mathrm th}$ transit. $\V$ is a function that can be calculated through the curve itself from the variance of the average flux over transit-length intervals outside the transit.

Consequently the significance (signal-to-noise) of a transit signal candidate in the presence of real photometric noise can be calculated by:
\begin{equation}
\Sr^2= d^2 \frac{n^2}{\sum_{k=1}^{N_{tr}} n_k^2 \V (n_k)} \ .
\end{equation}

Introducing the $\sigma_r$ factor to parametrize the red noise, the threshold for detection is:
\[
\frac{d^2 n^2}{ \sum_{k=1}^{N_{tr}} n_k^2 \left( \frac{\sigma_w^2}{n_k} + \sigma_r^2\right)} > S_T^2 \ ,
\]
where $\sigma_w$ is the uncorrelated (white) noise in the photometry. Typical values of $\sigma_r$ are of a few millimagnitudes (3 mmag in the OGLE survey), and typical values of $S_T$ are $9$ or higher in realistic settings.

In terms of physical parameters the threshold is:
\[
\Sr^2  = \alpha (r/R)^4 \left[ c \pi \frac{\sigma_w^2 P^{2/3} M^{1/3}}{\beta N R} + \frac{\sigma_r^2 P }{N \delta_t}\right]^{-1} > S_T^2 \ .
\]

Our approach bears similarities with that of \citet{jen02}, \citet{kov02} and \citet{sir03}, in accounting for the systematics in the noise by studying the properties of the photometric time series over longer time intervals. We tried to devise a method that is easy to apply, does not require long computation, and can be extrapolated to ongoing or future surveys with minimal extra assumptions.

In general, accounting for the effect of systematics results in much
lower yield estimates for ground-based transit surveys. We believe
that these estimates are more realistic than white-noise
estimates. They are also in much better agreement with the low actual
detection rate. These more realistic estimates can help in the design
of transit surveys. Including the red noise modifies the dependence of
the yield estimates on the survey parameters. Thus, survey design can
be affected not only quantitatively by lowering the predicted yields,
but also qualitatively.  The presence of systematics has several
important effects: (1) it reduces the premium placed on the brightest
end of a survey for detection, (2) it steepens the period dependence
of detectability and (3) it reduces the importance of denser time
sampling.

These elements may for instance weigh a survey strategy towards including more fields with a longer separation between individual measurements, and spending more time on a given field.

In the case of surveys targeting stellar systems, red noise modifies the conclusions of \citet{gau05} that if the detection threshold reaches a hot Jupiter transit in front of a target of any magnitude, such transits will be detected for targets of all magnitudes. Red noise will hamper the detection for the brightest magnitudes. 
 
Our study also underlines the crucial importance of beating down the
hour-timescale systematics in the photometry of transit surveys. This,
of course, is well realized by all groups conducting such surveys, and
considerable effort has been invested in removing the systematics. For
instance, for the OGLE survey, two schemes have been developed to
remove the systematics by using the correlation among all the
lightcurves \citep{kru03,tam05}.

However, for a given facility there may be a lower limit below which
it will be practically impossible to reduce $\sigma_r$. Since transit
surveys need as large a time coverage as they can get, they are
compelled to use all clear nights whether perfectly photometric or
not. At the millimagnitude level, interactions between airmass,
colour, absorption, tracking, flatfielding and seeing all have the
potential to introduce systematic trends. For instance,
systematic residuals can be caused by the interaction of seeing with a
close unseen companion, or by the tracking drift moving part of an
object across pixels with flatfield errors.

Sub-millimagnitude accuracy has been reported in some occasions, for instance by \citet{har05}. This can be attained with a very good sampling of the PSF on a large number of pixels, as possible with a large telescope on small fields. However, the reported results concern only a few nights, probably of above-average quality. It remains to be seen if such accuracy can be maintained over a period of time long enough  for an efficient planetary transit survey. Moreover, the use of a large telescope on a small field implies that the detected transit candidates will be too faint for spectroscopic confirmation, limiting their usefulness. As shown by the existing surveys, most transit candidates are actually eclipsing binaries, so that spectroscopic follow-up is indispensable in order to confirm the planetary nature of the transiting companion.

An obvious solution to the problem of photometric
systematics is to move into space, as was indeed done soon after the
discovery of 51\,Peg with {\it HST} in the globular cluster 47\,Tuc
\citep{gil00}. More recently, Sahu et al. (programme GO-9750) have carried
out an HST survey for transits in the Galactic plane. The problem,
however, is that the detected transit candidates are again too faint
for spectroscopic follow-up, so that planets cannot be told apart from
stellar eclipses.

In the coming years, two space-born planetary transit missions are
scheduled, {\it CoRoT} and {\it Kepler}. They will be free from
systematics caused by the atmosphere. The approach of this paper,
however, also applies to space-born data. The transit detection
threshold for {\it CoRoT} and {\it Kepler} will also be set by the
characteristics of the red noise, both instrumental and due to stellar
variability, because transit-length intervals will be sufficient for
the white noise to average out.  The most relevant factor will be the
stability of
the hour-timescale average of the measured fluxes.

In summary, this work offers a practical way, both conceptually and
computationally, to empirically deal with the complex problem of
covariant noise in extrasolar planetary transit surveys, both 
ground-based and space-born.


\bibliography{pont_zucker}

\label{lastpage}

\end{document}